\documentclass[zpreprint,zbstepj]{./LaTeX/zeus/zeus_paper}
\usepackage[english]{babel}

\newcommand{\ZcoosysB}{%
The ZEUS coordinate system is a right-handed Cartesian system, with the $Z$
axis pointing in the proton beam direction, referred to as the ``forward
direction'', and the $X$ axis pointing left towards the centre of HERA.
The coordinate origin is at the nominal interaction point.\xspace}

\newcommand{\ZcoosysfnB}{\footnote{\ZcoosysB}}

\newcommand{\Zctddesc}[1]{%
Charged particles are tracked in the central tracking detector (CTD)~\citeCTD,
which operates in a magnetic field of $1.43\Tesla$ provided by a thin 
superconducting solenoid. The CTD consists of 72~cylindrical drift chamber 
layers, organized in nine superlayers covering the polar-angle#1 region 
\mbox{$15^\circ<\theta<164^\circ$}. The transverse-momentum resolution for
full-length tracks is $\sigma(p_T)/p_T=0.0058p_T\oplus0.0065\oplus0.0014/p_T$,
with $p_T$ in $\Gev$.}
\newcommand{\Zcaldesc}{%
The high-resolution uranium--scintillator calorimeter (CAL)~\citeCAL consists 
of three parts: the forward (FCAL), the barrel (BCAL) and the rear (RCAL)
calorimeters. Each part is subdivided transversely into towers and
longitudinally into one electromagnetic section (EMC) and either one (in RCAL)
or two (in BCAL and FCAL) hadronic sections (HAC). The smallest subdivision of
the calorimeter is called a cell.  The CAL energy resolutions, as measured under
test-beam conditions, are $\sigma(E)/E=0.18/\sqrt{E}$ for electrons and
$\sigma(E)/E=0.35/\sqrt{E}$ for hadrons, with $E$ in $\Gev$.}

\chardef\usc=95
\chardef\til=126
\catcode`\@=11 
\DeclareRobustCommand\xdotspace{\futurelet\@let@token\@xdotspace}
\def\@xdotspace{%
  \ifx\@let@token.\else
  \ifx\@let@token\bgroup.\else
  \ifx\@let@token\egroup.\else
  \ifx\@let@token\/.\else
  \ifx\@let@token\ .\else
  \ifx\@let@token~.\else
  \ifx\@let@token!.\else
  \ifx\@let@token,.\else
  \ifx\@let@token:.\else
  \ifx\@let@token;.\else
  \ifx\@let@token?.\else
  \ifx\@let@token/.\else
  \ifx\@let@token'.\else
  \ifx\@let@token).\else
  \ifx\@let@token-.\else
  \ifx\@let@token\@xobeysp.\else
  \ifx\@let@token\space.\else
  \ifx\@let@token\@sptoken.\else
   .\space
   \fi\fi\fi\fi\fi\fi\fi\fi\fi\fi\fi\fi\fi\fi\fi\fi\fi\fi}
\catcode`\@=12 

\newcommand{\stru}[2]{%
   \relax\ifmmode\hbox{\vrule height#1 depth#2 width0pt}%
   \else\vrule height#1 depth#2 width0pt\fi}

\newcommand{\Ronum}[1]{\uppercase\expandafter{\romannumeral#1}}
\newcommand{\ronum}[1]{\expandafter{\romannumeral#1}}
\DeclareRobustCommand{\LaTeXZ}{%
  \LaTeX\kern-.05em4\kern-.1em
  {\raisebox{-0.2ex}{$\scriptstyle\text{ZEUS}$}}\xspace}

\DeclareMathAlphabet{\mathbf}{OT1}{cmr}{bx}{sl}
\newcommand{\eVdist}{\kern-0.06667em}

\newcommand{\Gev}{{\text{Ge}\eVdist\text{V\/}}}

\newcommand{\gev}{{\,\text{Ge}\eVdist\text{V\/}}}

\newcommand{\pbi}{\,\text{pb}^{-1}}

\newcommand{\met}{\,\text{m}}

\newcommand{\Tesla}{\,\text{T}}

\newcommand{\slashfrac}[2]{%
  \raisebox{0.5ex}{\ensuremath #1}\kern-0.12em/\kern-0.08em
  \raisebox{-.8ex}{\ensuremath #2}}

\newcommand{\sqr}[3]{%
    {\vcenter{\hrule height.#3ex\hbox{\vrule width.#2ex height#1ex
     \kern#1ex\vrule width.#3ex}\hrule height.#2ex}}}

\catcode`\@=11 
\newcommand{\parenbar}{\mathpalette\p@renb@r}
\def\p@renb@r#1#2{\vbox{%
  \ifx#1\scriptscriptstyle \dimen@.7em\dimen@ii.2em\else
  \ifx#1\scriptstyle \dimen@.8em\dimen@ii.25em\else
  \dimen@1em\dimen@ii.4em\fi\fi \offinterlineskip
  \ialign{\hfill##\hfill\cr
    \vbox{\hrule width\dimen@ii}\cr
    \noalign{\vskip-.3ex}%
    \hbox to\dimen@{$\mathchar300\hfil\mathchar301$}\cr
    \noalign{\vskip-.3ex}%
    $#1#2$\cr}}}
\catcode`\@=12 

\newcommand{\IP}{{\rm I$\kern-0.01667em$P}\xspace}

\mathchardef\qsm=63
\mathchardef\pls=43
\mathchardef\mns=512
\mathchardef\plm=518
\mathchardef\eql=61
\mathchardef\smallleft=300
\mathchardef\smallright=301
\mathchardef\les=316
\mathchardef\gre=318
\mathchardef\leq=532
\mathchardef\grq=533

\catcode`\@=11 
\newcounter{pict@width}
\newcounter{pict@height}
\newlength{\pict@scale}
\newcommand{\psfigadd}[4]{%
\setcounter{pict@width}{1*\ratio{#2+\pict@scale/2}{\pict@scale}}
\setcounter{pict@height}{1*\ratio{#3+\pict@scale/2}{\pict@scale}}
\setlength{\unitlength}{\pict@scale}
\hbox to #2{\hspace{-\fill}\begin{picture}(\thepict@width,\thepict@height)
\put(0,0){\psfig{figure=#1,width=#2,height=#3,clip=}}
\SetScale{0.283466457}
\SetWidth{1.763889}
{#4}
\end{picture}}
}
\newcounter{pict@widthfst}
\newcounter{pict@widthscd}
\newcounter{pict@widthtot}
\newcommand{\psfigaddtwo}[7]{%
\setcounter{pict@widthfst}{1*\ratio{#2+\pict@scale/2}{\pict@scale}}
\setcounter{pict@widthscd}{1*\ratio{#2+#4+\pict@scale/2}{\pict@scale}}
\setcounter{pict@widthtot}{1*\ratio{#2+#4+#6+\pict@scale/2}{\pict@scale}}
\setcounter{pict@height}{1*\ratio{#3+\pict@scale/2}{\pict@scale}}
\setlength{\unitlength}{\pict@scale}
\hbox{\hspace{-\fill}\begin{picture}(\thepict@widthtot,\thepict@height)
\put(0,0){\psfig{figure=#1,width=#2,height=#3,clip=}}
\put(\thepict@widthscd,0){\psfig{figure=#5,width=#6,height=#3,clip=}}
\SetScale{0.283466457}
\SetWidth{1.763889}
{#7}
\end{picture}}
}
\newcommand{\psfigror}[4]{%
\setcounter{pict@width}{1*\ratio{#2+\pict@scale/2}{\pict@scale}}
\setcounter{pict@height}{1*\ratio{#3+\pict@scale/2}{\pict@scale}}
\setlength{\unitlength}{\pict@scale}
\hbox{\begin{picture}(\thepict@width,\thepict@height)
\put(0,\thepict@height){\psfig{figure=#1,width=#3,height=#2,clip=,angle=270}}
\SetScale{0.283466457}
\SetWidth{1.763889}
{#4}
\end{picture}}
}
\newcommand{\psfigrol}[4]{%
\setcounter{pict@width}{1*\ratio{#2+\pict@scale/2}{\pict@scale}}
\setcounter{pict@height}{1*\ratio{#3+\pict@scale/2}{\pict@scale}}
\setlength{\unitlength}{\pict@scale}
\hbox{\begin{picture}(\thepict@width,\thepict@height)
\put(0,0){\psfig{figure=#1,width=#3,height=#2,clip=,angle=90}}
\SetScale{0.283466457}
\SetWidth{1.763889}
{#4}
\end{picture}}
}
\catcode`\@=12 
\newlength\listtextwidth

\catcode`\@=11 
\newlength{\@tabfninsert}
\newlength{\@tabfnwidth}
\newcommand{\tabfootnote}[2]{%
  \setlength{\@tabfninsert}{0.8em}
  \setlength{\@tabfnwidth}{\textwidth}
  \addtolength{\@tabfnwidth}{-\@tabfninsert}
  \addtolength{\@tabfnwidth}{-0.4em}
  \noindent\makebox[\@tabfninsert][r]{\footnotesize$^{#1}$\hfil}\hfill%
  \parbox[t]{\@tabfnwidth}{\footnotesize #2\hfill}}
\catcode`\@=12 

\newcommand{\dsp}        {\mbox{$D^{\ast +}$}}
\newcommand{\dspm}       {\mbox{$D^{\ast \pm}$}}
\newcommand{\dz}         {\mbox{$D^{0}$}}
\newcommand{\done}       {\mbox{$D_1^{0}$}}
\newcommand{\dtwo}       {\mbox{$D_2^{\ast 0}$}}

\newcommand{\fcds}       {\mbox{$f(c \rightarrow D^{\ast +})$}}
\def\dsk3pi{ {\dsp}~\rightarrow~\dz~\pi^{+}_{s}%
        \rightarrow~(K^{-}~\pi^{+}~\pi^{+}~\pi^{-})~\pi^{+}_{s} }
\def\et10t{ E_T^{\theta > 10^\circ}}
\def\etw10{ E_T^{\theta > 10}}
\newcommand{\ra}         {\mbox{$\rightarrow$}}
\def\citeCTD{{\cite{%
nim:a279:290,*npps:b32:181,*nim:a338:254%
}}\xspace}
\def\citeCAL{{\cite{%
nim:a309:77,*nim:a309:101,*nim:a321:356,*nim:a336:23%
}}\xspace}

\begin{document}
\prepnum{{DESY--04--164}}

\title{
Search for a narrow charmed baryonic state decaying to $\dspm p^\mp$
in $ep$ collisions at HERA
}                                                       
                    
\author{ZEUS Collaboration}

\date{September 2004}

\abstract{
A resonance search has been made in the $\dspm p^\mp$ invariant-mass
spectrum with the ZEUS detector at HERA using an 
integrated luminosity of $126\pbi$.
The decay channels
$\dsp\rightarrow\dz\pi^{+}_{s}\rightarrow(K^{-}\pi^{+})\pi^{+}_{s}$
and
$\dsp\rightarrow\dz\pi^{+}_{s}\rightarrow(K^{-}\pi^{+}\pi^{+}\pi^{-})\pi^{+}_{s}$
(and the corresponding antiparticle decays) 
were used to identify $\dspm$ mesons.
No resonance structure was observed in the $\dspm p^\mp$ mass spectrum
from more than $60\,000$ reconstructed $\dspm$ mesons.
The results are not compatible with
a report of the H1 Collaboration of
a charmed pentaquark, $\Theta^0_c$.
}

\makezeustitle

\def\3{\ss}                                                                                        
\pagenumbering{Roman}                                                                              
                                                   %
\begin{center}                                                                                     
{                      \Large  The ZEUS Collaboration              }                               
\end{center}                                                                                       
  S.~Chekanov,                                                                                     
  M.~Derrick,                                                                                      
  J.H.~Loizides$^{   1}$,                                                                          
  S.~Magill,                                                                                       
  S.~Miglioranzi$^{   1}$,                                                                         
  B.~Musgrave,                                                                                     
  \mbox{J.~Repond},                                                                                
  R.~Yoshida\\                                                                                     
 {\it Argonne National Laboratory, Argonne, Illinois 60439-4815}, USA~$^{n}$                       
\par \filbreak                                                                                     
  M.C.K.~Mattingly \\                                                                              
 {\it Andrews University, Berrien Springs, Michigan 49104-0380}, USA                               
\par \filbreak                                                                                     
  N.~Pavel \\                                                                                      
  {\it Institut f\"ur Physik der Humboldt-Universit\"at zu Berlin,                                 
           Berlin, Germany}
% ...
\newpage                                                                                           
$^{\    1}$ also affiliated with University College London, UK \\                                  
% ...
                                                           %
                                                           %
\newpage   
                                                           %
                                                           %
\begin{tabular}[h]{rp{14cm}}                                                                       
$^{a}$ &  supported by the Natural Sciences and Engineering Research Council of Canada (NSERC) \\  
$^{b}$ &  supported by the German Federal Ministry for Education and Research (BMBF), under        
          contract numbers HZ1GUA 2, HZ1GUB 0, HZ1PDA 5, HZ1VFA 5\\                                
$^{c}$ &  supported in part by the MINERVA Gesellschaft f\"ur Forschung GmbH, the Israel Science   
          Foundation (grant no. 293/02-11.2), the U.S.-Israel Binational Science Foundation and    
          the Benozyio Center for High Energy Physics\\                                            
$^{d}$ &  supported by the German-Israeli Foundation and the Israel Science Foundation\\           
$^{e}$ &  supported by the Italian National Institute for Nuclear Physics (INFN) \\                
$^{f}$ &  supported by the Japanese Ministry of Education, Culture, Sports, Science and Technology 
          (MEXT) and its grants for Scientific Research\\                                          
$^{g}$ &  supported by the Korean Ministry of Education and Korea Science and Engineering          
          Foundation\\                                                                             
$^{h}$ &  supported by the Netherlands Foundation for Research on Matter (FOM)\\                   
$^{i}$ &  supported by the Polish State Committee for Scientific Research, grant no.               
          620/E-77/SPB/DESY/P-03/DZ 117/2003-2005 and grant no. 1P03B07427/2004-2006\\             
$^{j}$ &  partially supported by the German Federal Ministry for Education and Research (BMBF)\\   
$^{k}$ &  supported by RF President grant N 1685.2003.2 for the leading scientific schools and by  
          the Russian Ministry of Industry, Science and Technology through its grant for           
          Scientific Research on High Energy Physics\\                                             
$^{l}$ &  supported by the Spanish Ministry of Education and Science through funds provided by     
          CICYT\\                                                                                  
$^{m}$ &  supported by the Particle Physics and Astronomy Research Council, UK\\                   
$^{n}$ &  supported by the US Department of Energy\\                                               
$^{o}$ &  supported by the US National Science Foundation\\                                        
$^{p}$ &  supported by the Polish Ministry of Scientific Research and Information Technology,      
          grant no. 112/E-356/SPUB/DESY/P-03/DZ 116/2003-2005 and 1 P03B 065 27\\                  
$^{q}$ &  supported by FNRS and its associated funds (IISN and FRIA) and by an Inter-University    
          Attraction Poles Programme subsidised by the Belgian Federal Science Policy Office\\     
\end{tabular}                                                                                      
                                                           %
                                                           %

\pagenumbering{arabic} 
\pagestyle{plain}
\section{Introduction}
\label{sec-int}

The observation of a narrow strange baryonic state decaying to $K^+n$ or
$K^0_s p$ has been reported by several
experiments~\cite{prl:91:012002,*pan:66:1715,*prl:91:252001,*pl:b572:127,*pan:67:682,*prl:92:032001,*pl:b585:213,*hep-ex-0401024,*pl:b595:127}.
This state has both baryon number and strangeness of $+1$. Thus
the resonance cannot be composed of three quarks but could be explained
as a bound state
of five quarks: $\Theta^+ = uudd{\bar s}$.
Evidence for two other pentaquark states with strangeness of $-2$
has also been reported recently~\cite{prl:92:042003}.
Although no strange pentaquark production has been observed
in some searches~\cite{pr:d70:012004,*herab},
the existence of $\Theta^+$ is supported by recent results obtained
in $ep$ collisions at HERA~\cite{pl:b591:7}.
Several QCD models are able to explain
the nature of the strange pentaquarks~\cite{zfp:a359:305,prl:91:232003,etf:97:433,*pl:b570:185,*pl:b575:249}.

The expected properties of charmed pentaquark states have been
discussed in the literature~\cite{pl:b193:323,*pl:b195:484,*pl:b299:338,*pr:d58:111501,prl:91:232003,hep-ph-0307343,pr:d69:094029,*hep-ph-0402244,*hep-ph-0403232,*hep-ph-0403235}.
The lightest charmed pentaquark
would be $\Theta^0_c = uudd{\bar c}$.
In predictions based on the diquark-diquark-antiquark model
of Jaffe and Wilzcek~\cite{prl:91:232003},
the mass of $\Theta^0_c$ is typically below the sum of the masses of
the $D^-$ meson
and proton.
In this case, $\Theta^0_c$ should
decay weakly to, e.g., $\Theta^+ \pi^-$.
Predictions utilising the diquark-triquark model of
Karliner and Lipkin~\cite{hep-ph-0307343} suggest a heavier
$\Theta^0_c$ decaying
dominantly to $D^-p$.
If the mass of the $\Theta^0_c$ were sufficiently large,
it could also decay to $D^{*-}p$;
there is a possibility that this decay mode
is dominant~\cite{misc:karliner:private}.

The observation of a narrow charmed baryonic resonance decaying to
$\dspm p^\mp$ has recently been reported by
the H1 Collaboration~\cite{pl:b588:17}.
A peak containing $50.6\pm11.2$ events was observed in deep inelastic
scattering (DIS) with exchanged photon virtuality
$Q^2>1\gev^2$
at a mass of $3099\pm3({\rm stat.})\pm5({\rm syst.})\,$MeV
and with a Gaussian width of $12\pm3({\rm stat.})\,$MeV,
compatible with the experimental resolution.
A signal with compatible mass and width
was also observed in photoproduction ($Q^2\lesssim 1\gev^2$).
The observed resonance was claimed to contribute roughly $1\%$
to the total $\dspm$ production rate
in the kinematic region studied.

This paper presents results of a search for narrow states in
the $\dspm p^\mp$ decay channel in
$e^\pm p$ collisions at HERA using the ZEUS detector.

\section{Experimental set-up}
\label{sec:expset}

The analysis was performed with the data taken by the ZEUS Collaboration
from 1995 to 2000.
In this period, HERA collided electrons or positrons\footnote{
From now on, the word ``electron'' is used as a generic term
for electrons and positrons.}
with energy $E_e=27.5\gev$ and protons with energy $E_p=820\gev$ (1995-1997)
or $E_p=920\gev$ (1998-2000).
The data used in this analysis correspond to an integrated luminosity
of $126.5\pm2.4\pbi$.

A detailed description of the ZEUS detector can be found 
elsewhere~\cite{zeus:1993:bluebook}. A brief outline of the 
components most relevant to this analysis is given
below.

\Zctddesc\ZcoosysfnB
~To estimate the energy loss per unit length, $dE/dx$, of particles in
the CTD~\cite{pl:b481:213,*epj:c18:625},
the truncated mean of the anode-wire pulse heights was calculated,
which
removes the lowest $10\%$ and at least the highest $30\%$
depending on the number of saturated hits.
The measured $dE/dx$ values were corrected by normalising to
the $dE/dx$ peak position for tracks around the region of minimum ionisation
for pions, $0.3<p<0.4\,$GeV. Henceforth $dE/dx$ is quoted in units of minimum
ionising particles (mips).
The resolution of the $dE/dx$ measurement
for full-length tracks is about $9\%$.

\Zcaldesc
~The position of electrons scattered with a small angle with respect to the 
electron beam direction was measured using the small-angle rear tracking 
detector (SRTD)~\cite{nim:a401:63}.

The luminosity was determined from the rate of the bremsstrahlung process
$ep \rightarrow e \gamma p$, where the photon was measured with a 
lead--scintillator calorimeter~\cite{desy-92-066,*zfp:c63:391,*acpp:b32:2025} 
located at $Z = -107\met$.

\section{Event simulation}
\label{sec-simul}

Monte Carlo (MC) samples of charm and beauty events
were produced with
the {\sc Pythia} 6.156~\cite{cpc:82:74} and
{\sc Rapgap}~2.0818~\cite{cpc:86:147}
event generators.
The generation included direct photon processes,
in which the photon couples directly to a parton in the proton,
and resolved photon processes, where the photon acts as a source
of partons, one of which participates in the hard scattering process.
The CTEQ5L~\cite{epj:c12:375} and GRV~LO~\cite{pr:d46:1973} parameterisations
were used for the proton and photon structure functions, respectively.
The Lund string model~\cite{prep:97:31}
as implemented in {\sc Jetset}~\cite{cpc:82:74}
was used for hadronisation.
The Bowler modification~\cite{zfp:c11:169}
of the LUND symmetric fragmentation function~\cite{zfp:c20:317}
was used for the charm and bottom quark fragmentation.
The charm and bottom quark masses were set to the values $1.5\,$GeV and
$4.75\,$GeV, respectively.
All processes were generated in proportion to the predicted MC cross
sections.
The combined sample of the {\sc Pythia} events, generated with $Q^2<0.6\gev^2$,
and the {\sc Rapgap} events, generated with $Q^2>0.6\gev^2$,
was used as the inclusive $D^{*\pm}$ MC sample
after reweighting the
$\dspm$ transverse momentum,
$p_T(D^{* \pm})$, and pseudorapidity, $\eta(D^{* \pm})$, distributions
to describe the data.

To generate the $\Theta^0_c$, the mass
of a neutral charmed baryon
in the {\sc Jetset} particle table
was set to $3.099\,$GeV~\cite{pl:b588:17},
its width was set to zero and the decay channel was set
to $D^{*-}p$.
The $\Theta^0_c$ samples produced with the {\sc Pythia} and {\sc Rapgap}
generators
were combined in the same way as
described in the previous paragraph.
Since the production mechanism of the $\Theta^0_c$ is unknown,
the simulated
$p_T(\Theta^0_c)$ and $\eta(\Theta^0_c)$ distributions
were reweighted to the
$p_T(D^{* \pm})$ and $\eta(D^{* \pm})$
distributions of the inclusive $\dspm$ MC which describes the data.

The generated events were passed through a full simulation
of the detector using {\sc Geant} 3.13~\cite{tech:cern-dd-ee-84-1}
and processed with the same reconstruction program as used for the data.

\section{Event selection and reconstruction of $\dspm$ mesons}
\label{sec-seldstar}

The $\dspm(2010)$ mesons were identified using the two decay channels

\begin{equation}
\dsp\rightarrow\dz\pi^{+}_{s}\rightarrow(K^{-}\pi^{+})\pi^{+}_{s},
\end{equation}
\begin{equation}
\dsp\rightarrow\dz\pi^{+}_{s}\rightarrow(K^{-}\pi^{+}\pi^{+}\pi^{-})\pi^{+}_{s}.
\end{equation}

Charge-conjugate processes are included.
The $\pi_s$ particle from the $\dspm$ decay is referred to as the ``soft pion''
because it is constrained to have limited momentum
by the small mass difference between the $\dsp$ and $\dz$.

Events from both photoproduction~\cite{epj:c6:67} and DIS~\cite{epj:c12:35,*pr:d69:0120004}
were selected online with a three-level
trigger~\cite{pl:b293:465,zeus:1993:bluebook}.
At the third level,
where the full event information was available,
the nominal $D^*$ trigger branch required the
presence of a reconstructed
$D^*$-meson candidate and, for DIS, a scattered-electron candidate.
The efficiency of the online $D^*$ reconstruction, determined
relative to the efficiency of the offline $D^*$ reconstruction using
an inclusive DIS trigger and a photoproduction dijet trigger,
was above $95\%$
for most of the data-taking period.
Events missed by the nominal $D^*$ trigger but selected with some other
trigger branch were also used in this analysis.

In the offline analysis,
only events with $|Z_{\rm vertex}|<50\,$cm,
where $Z_{\rm vertex}$ is the primary vertex position determined
from the CTD tracks,
were used.
For each event,
a search for
the scattered electron from the pattern
of energy deposits in the CAL~\cite{nim:a365:508}
was performed.
If a scattered-electron candidate was found,
the following criteria were imposed
to select DIS events:
\begin{itemize}
\item the scattered electron energy above $8\,$GeV;
\item the impact point of the scattered electron on the RCAL
outside the $(X,Y)$ region $(24{\rm cm},12{\rm cm})$ centered on the beamline;
\item $40<E-P_{Z}<65\,$GeV, where
$E-P_{Z}={\Sigma_{i}(E-P_{Z})_{i}}$ and
the sum runs over a combination of charged tracks,
as measured in the CTD,
and energy clusters measured in the CAL~\cite{briskinu:phd:1998};
\item $y<0.95$, where $y$ is the fraction of the electron energy
transferred to the proton in its rest frame.
For this cut, $y$ was calculated
from the energy and angle of
the scattered electron;
\item $Q^2>1\,$GeV$^2$, using measurements of the energy and angle of
the scattered electron.
\end{itemize}
All events which failed the DIS selection were assigned to the photoproduction
sample.
Monte Carlo studies showed
that $98\%$ of the DIS sample consisted of events
with true $Q^2>1\,$GeV$^2$
and $95\%$ of the photoproduction sample consisted of events
with true $Q^2<1\,$GeV$^2$.
The migrations were taken into account in the correction procedure
for detector effects
(Section 7).

In each event,
charged tracks measured by the CTD and assigned to the primary
event vertex were selected.
The transverse momentum was required to be greater than $0.1\,$GeV.
Each track was required to reach at least the third superlayer of the CTD.
These restrictions ensured both good track acceptance and
good momentum resolution.

Selected tracks were combined to form $\dz$ candidates
assuming the decay channels (1) or (2).
For both cases, $\dz$ candidates were formed by calculating the invariant
mass $M(K\pi)$ or $M(K\pi\pi\pi)$ for combinations having a total charge
of zero.
The soft pion was required to have
a charge opposite to that of the particle taken
as a kaon
and was used
to form a $D^*$ candidate
having mass $M(K \pi \pi_s)$ or $M(K \pi \pi \pi \pi_s)$.
No particle identification was used, so kaon and pion masses
were assigned in turn to each track.

To reduce the combinatorial background,
the following transverse momentum requirements were applied to tracks
from the above combinations:
$$p_T(K)>0.45\,{\rm GeV},\; p_T(\pi)>0.45\,{\rm GeV},\; p_T(\pi_s)>0.1\,{\rm GeV}$$
for channel (1), and
$$p_T(K)>0.5\,{\rm GeV},\; p_T(\pi)>0.2\,{\rm GeV},\; p_T(\pi_s)>0.15\,{\rm GeV}$$
for channel (2).
The $D^*$ candidates were required to have
$-1.6<\eta(D^*)<1.6$,
where the CTD acceptance is high.
Also,
$p_T(D^*)>1.35\gev$ or $p_T(D^*)>2.8\gev$ for channels (1) or (2),
respectively, was required
to further reduce the combinatorial background.

For selected $D^*$ candidates,
consistency of the $M(K \pi)$ or $M(K \pi \pi \pi)$ value
with the nominal $\dz$ mass was required.
To take account of the mass resolution, the requirement was
$$1.83<M(K \pi)<1.90\,{\rm GeV},\;1.845<M(K \pi \pi \pi)<1.885\,{\rm GeV}$$
for $p_T(D^*)<5\,{\rm GeV}$,
$$1.82<M(K \pi)<1.91\,{\rm GeV},\;1.835<M(K \pi \pi \pi)<1.895\,{\rm GeV}$$
for $5<p_T(D^*)<8\,{\rm GeV}$, and
$$1.81<M(K \pi)<1.92\,{\rm GeV},\;1.825<M(K \pi \pi \pi)<1.905\,{\rm GeV}$$
for $p_T(D^*)>8\,{\rm GeV}$.

To suppress the combinatorial background, a cut
on the ratio
$p_T(D^*)/\etw10$ was applied,
where $\etw10$ is the transverse energy measured in the
CAL outside a cone of $\theta=10^\circ$ around the
forward direction.
For DIS events, the energy assigned to the scattered electron was
excluded from the $\etw10$ calculation.
The cut value was $p_T(D^*)/\etw10>0.12$ and
$p_T(D^*)/\etw10>0.2$
for channels (1) and (2), respectively. Monte Carlo studies showed that
this cut removed a significant fraction
of the background whilst preserving most of the $D^*$ mesons produced
either inclusively or in $\Theta^0_c$ decays.

\begin{figure}[hbtp]
\epsfysize=18cm
\vspace*{-1.0cm}
\centerline{\epsffile{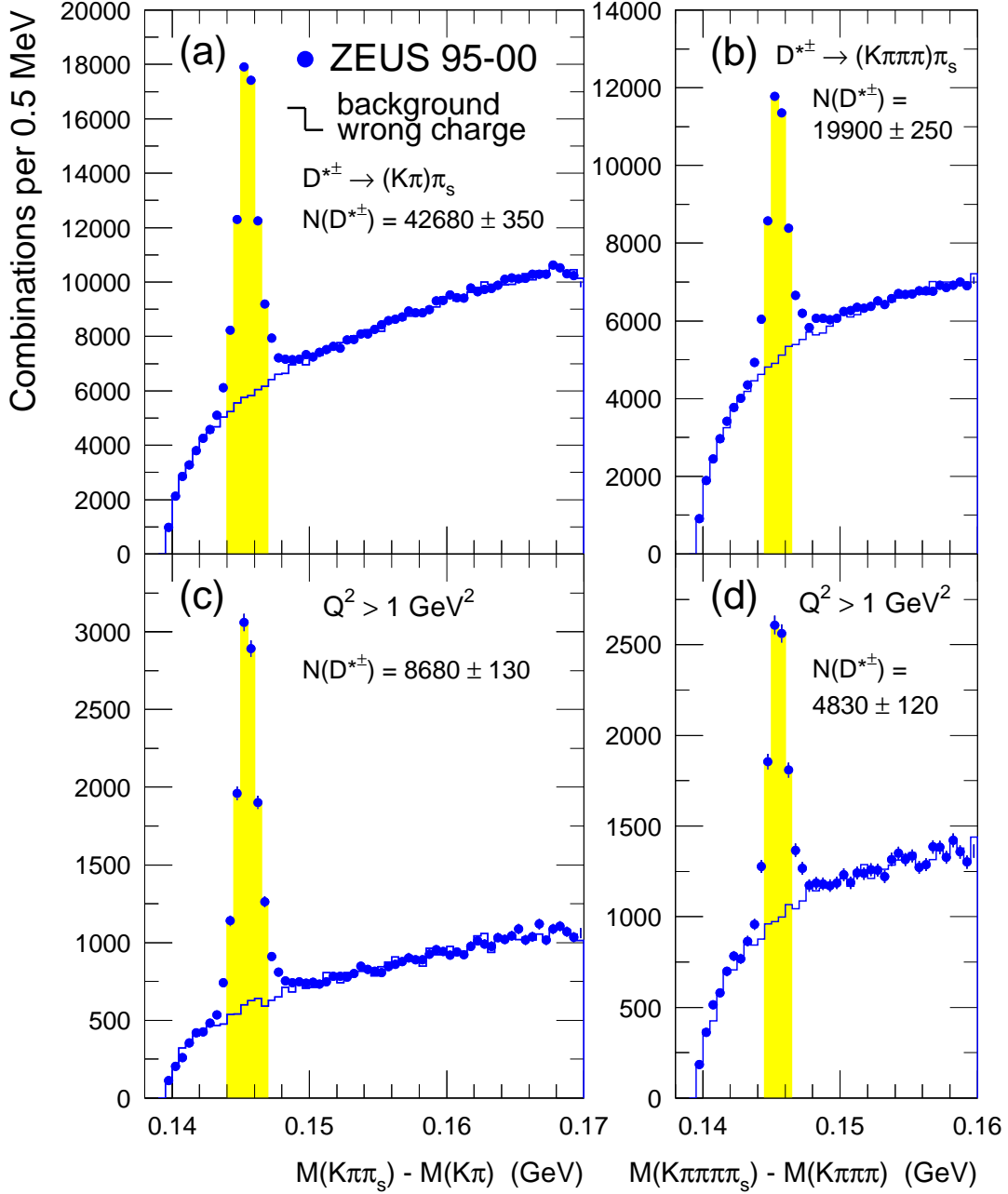}}
\caption{
The distribution of the mass difference,
$\Delta M$, (dots) for
(a) $D^{*\pm}\rightarrow (K\pi)\pi_s$ candidates in
the full data sample,
(b) $D^{*\pm}\rightarrow (K\pi\pi\pi)\pi_s$ candidates in
the full data sample,
(c) $D^{*\pm}\rightarrow (K\pi)\pi_s$ candidates in
DIS with $Q^2>1\gev^2$ and
(d) $D^{*\pm}\rightarrow (K\pi\pi\pi)\pi_s$ candidates in
DIS with $Q^2>1\gev^2$.
The histograms
show the $\Delta M$ distributions for wrong charge combinations.
Only $\dspm$ candidates from the shaded bands
were used for the charmed pentaquark search.
}
\label{fig:kpi_k3pi_dm}
\end{figure}

The mass difference $\Delta M=M(K \pi \pi_s)-M(K \pi)$ for channel (1) or
$\Delta M=M(K \pi \pi \pi \pi_s)-M(K \pi \pi \pi)$ for channel (2)
was evaluated for all remaining $D^*$ candidates.
Figures~\ref{fig:kpi_k3pi_dm}a and~\ref{fig:kpi_k3pi_dm}b
show the mass-difference
distributions for channels (1) and (2), respectively.
In Figs.~\ref{fig:kpi_k3pi_dm}c and~\ref{fig:kpi_k3pi_dm}d,
the mass-difference distributions
are shown for DIS events with $Q^2>1\,$GeV$^2$.
Peaks at the nominal value of $M(D^{*+})-M(D^0)$ are evident.
For channel (2), the same tracks can produce two $D^0$ candidates
due to an ambiguity in the kaon and pion mass assignment to tracks
with the same charge. Such candidates produce different $M(K \pi \pi \pi)$
values and almost identical $\Delta M$ values. To exclude double
counting, both combinations of the same tracks which passed all cuts,
including the $M(K \pi \pi \pi)$ requirement,
were included
with a weight $1/2$.

To determine the background under the peak, wrong-charge combinations
were used. For both channels (1) and (2), these
are defined as combinations with
total charge $\pm2$ for the $D^0$ candidate and total charge
$\pm1$ for the $D^*$ candidate.
The histograms in Fig.~\ref{fig:kpi_k3pi_dm} show
the $\Delta M$ distributions for the wrong-charge combinations,
normalised to the distributions of $D^*$ candidates with the appropriate
charges in the range $0.15<\Delta M<0.17\gev$ for channel (1) and
$0.15<\Delta M<0.16\gev$ for channel (2).
The upper ends of the normalisation ranges correspond to
the trigger selections of $D^*$ candidates in the two decay channels.
For both channels, the same tracks from a wrong-charge combination
can produce two $D^0$ candidates 
due to an ambiguity in the kaon and pion mass assignment to tracks
with the same charge. For channel (2), it is also possible to have three
wrong-charge
$D^0$ candidates produced by the same tracks.
To exclude double and triple
counting, the multiple combinations of the same tracks which passed all cuts,
including the $M(K \pi)$ or $M(K \pi \pi \pi)$ requirement,
were included
with a weight $1/2$ or $1/3$ for double or triple
entries, respectively.
Monte Carlo studies showed that the procedure
used for the background determination
and the treatment of multiple entries permits the recovery of the number
of true $D^*$ mesons for both channels (1) and (2).

To improve the signal-to-background ratio,
only $D^*$ candidates
with $0.144<\Delta M<0.147\gev$ for channel (1) and
$0.1445<\Delta M<0.1465\gev$ for channel (2) were kept
for the charmed pentaquark search.
After background subtraction,
signals of $42680\pm350$ $\dspm$ mesons in channel (1)
and $19900\pm250$ $D^*$ mesons in channel (2)
were found in the above $\Delta M$ ranges.
In DIS with $Q^2>1\,$GeV$^2$,
the numbers of reconstructed $D^*$ mesons were $8680\pm130$ in channel (1)
and $4830\pm120$ in channel (2), whereas for
$Q^2<1\,$GeV$^2$
$34000\pm330$ and $15070\pm220$ $D^*$ mesons 
were found
in channels (1) and (2),
respectively.

The relative acceptance for $D^*$ mesons originating
from the $\Theta^0_c$ and $D^*$ mesons produced inclusively,
$A^{\Theta^0_c}(D^*)/A^{\rm inc}(D^*)$, was calculated using
the $\Theta^0_c$ and the inclusive $D^*$ MC samples. The values
of this relative acceptance were $85\%$ and $87\%$ for
the samples with $D^*$ reconstructed in the decay channels (1) and (2),
respectively.

\section{Selection of proton candidates and $D^*p$ invariant mass reconstruction}
\label{sec-selprot}

A charmed pentaquark candidate
was formed by combining a selected $D^*$
candidate with a track,
assumed to be a proton, with $p_T > 0.15\,$GeV and a charge opposite
to that of the $D^*$.
For each charmed pentaquark candidate,
the ``extended" mass difference,
$\Delta M^{\rm ext} = M(K \pi \pi_s p)-M(K \pi \pi_s)$ or
$\Delta M^{\rm ext} = M(K \pi \pi \pi \pi_s p)-M(K \pi \pi \pi \pi_s)$,
was calculated.
The invariant mass of the $D^*p$ system was calculated as
$M(D^*p)=\Delta M^{\rm ext}+M(\dsp)_{\rm PDG}$,
where $M(\dsp)_{\rm PDG}$ is the nominal $\dspm$ mass~\cite{pr:d55:10001}.
The resolution in $M(D^*p)$ for $M(D^*p)\sim3.1\,$GeV,
where a narrow signal was reported by the H1 Collaboration~\cite{pl:b588:17},
was estimated from MC simulations
to be $4\,$MeV.

\begin{figure}[hbtp]
\epsfysize=18cm
\vspace*{-1.0cm}
\centerline{\epsffile{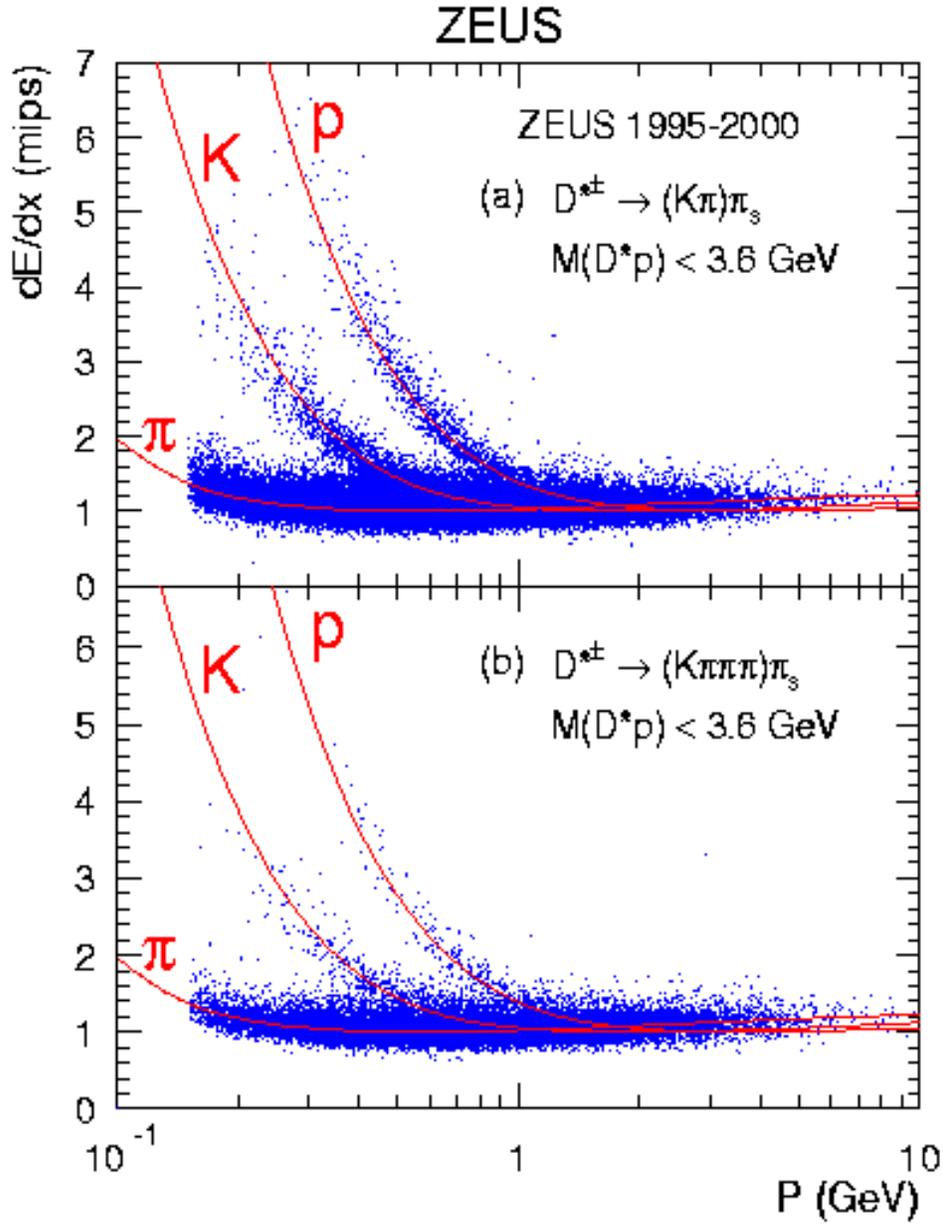}}
\caption{
The $dE/dx$ values as a function of momentum, $P$,
for particles which yield a mass $M(D^* p)<3.6\gev$
when combined with
(a) $D^{*\pm}\rightarrow (K\pi)\pi_s$ candidates and
(b) $D^{*\pm}\rightarrow (K\pi\pi\pi)\pi_s$ candidates.
The lines indicate parameterisations
for the expectation values of $dE/dx$
for pions, kaons and protons.
}
\label{fig:kpi_k3pi_dedx}
\end{figure}

To reduce pion and kaon backgrounds,
the measured $dE/dx$ values for proton candidates were used.
To ensure good $dE/dx$ resolution,
at least eight CTD hits were used.
Figure~\ref{fig:kpi_k3pi_dedx} shows
the $dE/dx$ values as a function of momentum, $P$,
for particles which yield a mass $M(D^* p)<3.6\,$GeV.
The proton bands are clearly seen in the distributions for
particles associated
with $D^*$ in both decay channels.
The parameterisation for the expectation value of $dE/dx$
as a function of $P/m$ was obtained
using tagged protons from $\Lambda^0$ decays
and tagged pions from $K^0_s$ decays.
To construct a $\chi^2_1$ function,
the following procedure~\cite{deppe:phd:1999} was used.
For each particle,
a $\chi^2_1$ value that estimates the deviation
of the measured $dE/dx$
from the expectation
was calculated as:
$$\chi^2_1=\frac{[\ln(dE/dx)-\ln(dE/dx)_{\rm expected}]^2}{\sigma^2_{\ln(dE/dx)}}.$$
The resolution was parameterised empirically as
$\sigma_{\ln(dE/dx)}=a/\sqrt{n}$, where $n$ is the number of hits used for
the $dE/dx$ measurement and $a$ is a constant determined from the sample of
tagged protons.
The $\chi^2_1$ probability of the proton hypothesis, $l_p$,
is given by
the probability for a proton to produce the observed or a larger
value of $\chi^2_1$.

The distribution of $l_p$ for proton candidates shows a sharp peak at
$l_p\sim 0$ and becomes relatively flat towards $l_p\sim 1$.
To maximise the ratio of the number of selected protons to the square root
of the number
of background particles, a cut
$l_p>0.15$ was applied.

The acceptance of the proton selection before the requirement on $l_p$,
$A(p)$,
was calculated using the $\Theta^0_c$ MC to be $85\%$ and $89\%$ for
the samples with $D^*$ reconstructed in the decay channels (1) and (2),
respectively.
The acceptance
$A(l_p>0.15)$
was calculated,
using the tagged protons, to be $(85.0\pm0.1)\%$. This acceptance,
calculated directly from the data, was insensitive to the proton
momenta spectrum.

\section{$D^*p$ invariant mass distributions}
\label{sec-distrib}

\begin{figure}[hbtp]
\epsfysize=18cm
\vspace*{-1.0cm}
\centerline{\epsffile{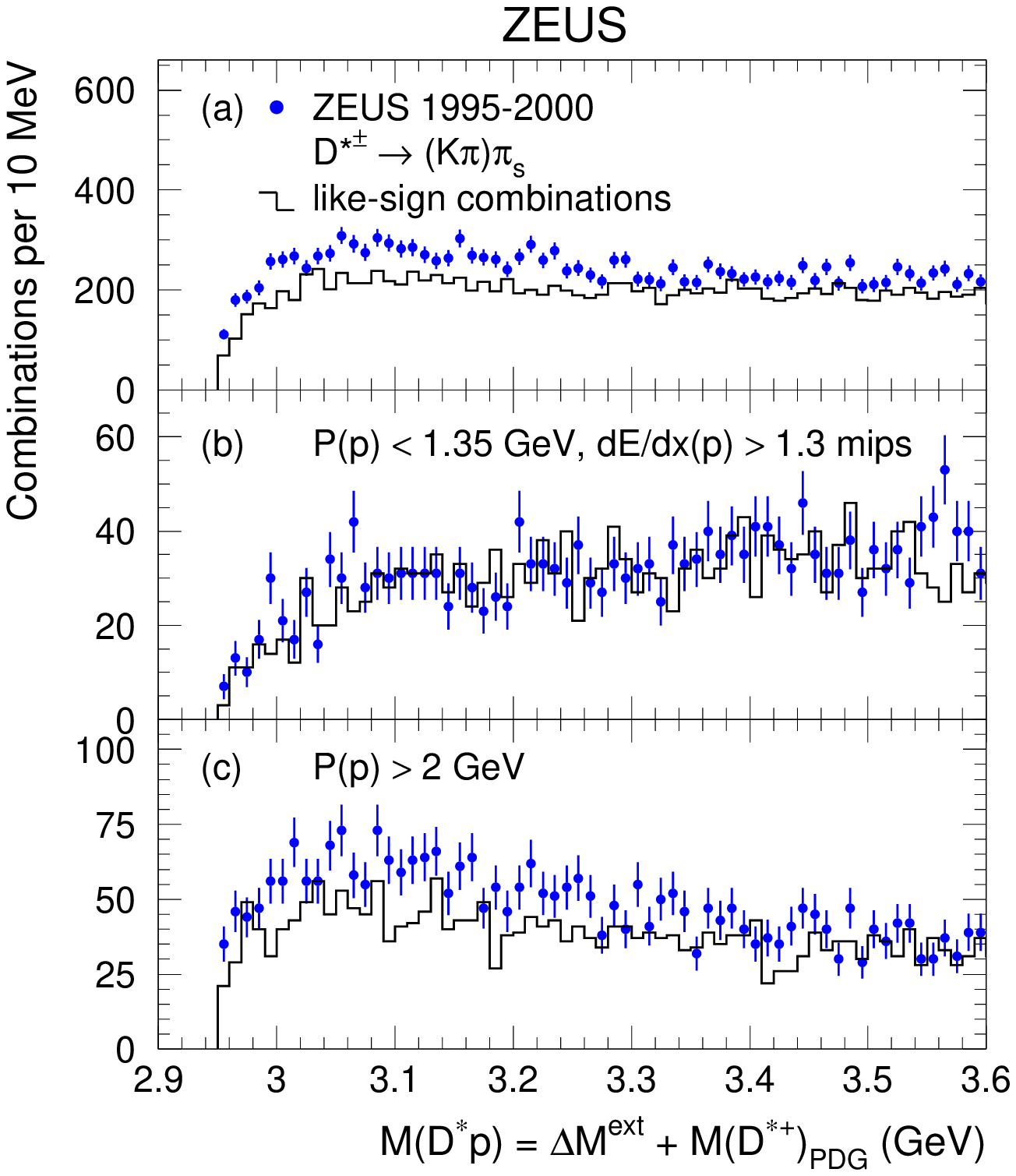}}
\caption{
The distribution of
$M(D^* p)=\Delta M^{\rm ext}+M(\dsp)_{\rm PDG}$
for charmed pentaquark candidates (dots)
obtained with the full data sample
using
(a) all proton candidates,
(b) proton candidates with momentum below $1.35\gev$ and $dE/dx$
above $1.3\,$mips, and
(c) proton candidates with momentum above $2\gev$.
The extended mass difference is defined as
$\Delta M^{\rm ext}=M(K \pi \pi_s p)-M(K \pi \pi_s)$ and
$M(\dsp)_{\rm PDG}$ is the nominal $\dsp$ mass.
The histograms  show the $M(D^* p)$
distributions for the like-sign combinations.
}
\label{fig:kpi_dp_all}
\end{figure}

Figure~\ref{fig:kpi_dp_all}a shows the $M(D^*p)$
distribution\footnote{
The $M(D^*p)$ distributions, shown in this paper
for $M(D^*p)<3.6\,$GeV,
were investigated in the full kinematically allowed range.}
for $D^*$ meson candidates reconstructed in the decay
channel (1).
No narrow resonance is seen.
To suppress the large background from pion and kaon tracks,
the following two selections were used in addition
to the general proton selection described in Section~5:
\begin{itemize}
\item low-momentum selection: only tracks with $P<1.35\,$GeV and
$dE/dx>1.3\,$mips were used as proton candidates. These requirements select
clean proton samples corresponding to the proton bands separated
from the pion and kaon bands
in Fig.~\ref{fig:kpi_k3pi_dedx};
\item high-momentum selection: only tracks with $P>2\,$GeV
were used as proton candidates. This selection was suggested by the
observation of the H1 Collaboration~\cite{pl:b588:17}
that the signal-to-background ratio
for charmed pentaquarks
improves as the proton momentum increases.
\end{itemize}
Figures~\ref{fig:kpi_dp_all}b and~\ref{fig:kpi_dp_all}c
show the $M(D^*p)$ distributions
for the low-momentum and high-momentum proton selections,
respectively.
The selections reveal no narrow resonance.

\begin{figure}[hbtp]
\epsfysize=18cm
\vspace*{-1.0cm}
\centerline{\epsffile{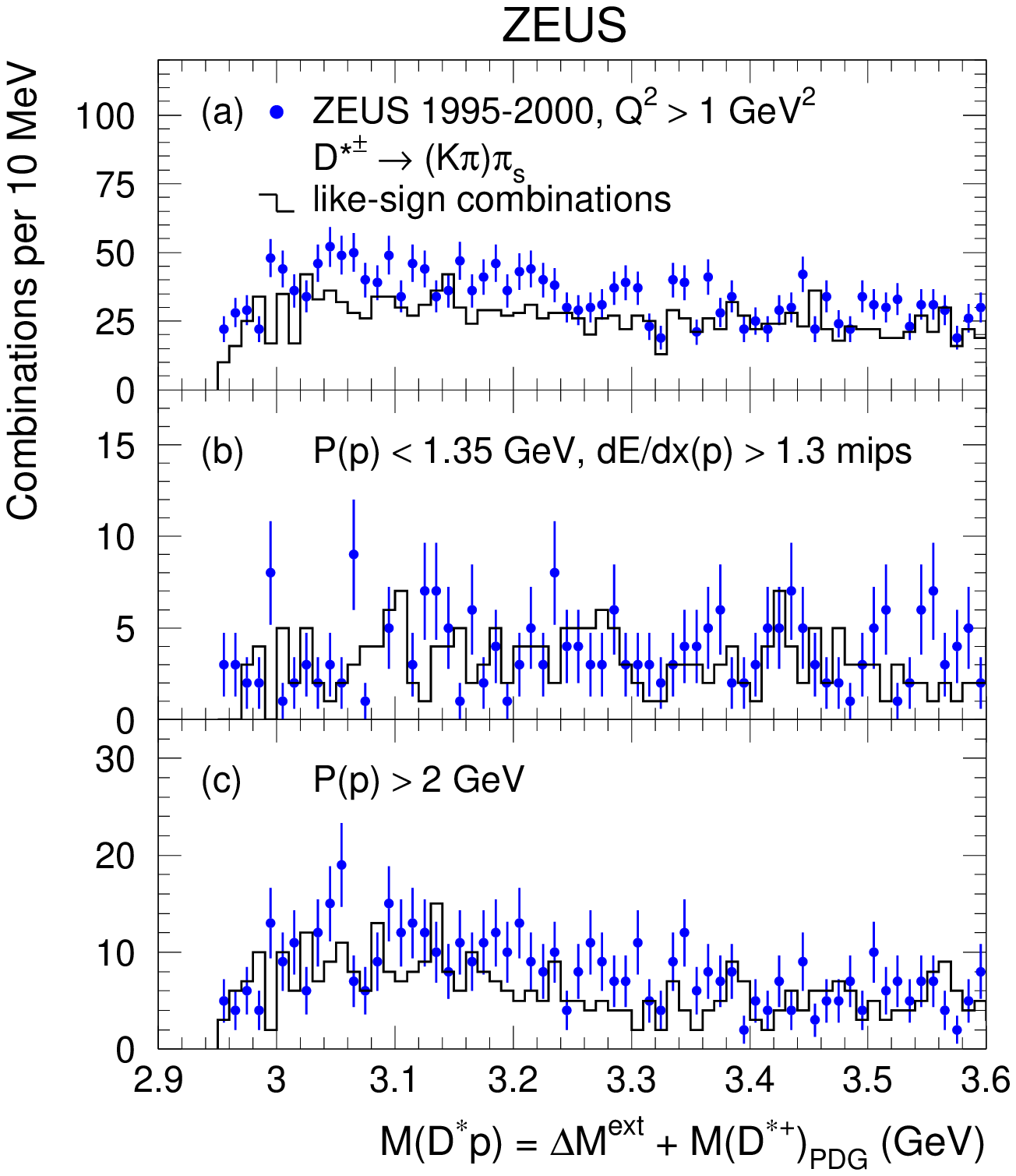}}
\caption{
The distribution of
$M(D^* p)=\Delta M^{\rm ext}+M(\dsp)_{\rm PDG}$
for charmed pentaquark candidates (dots)
obtained in DIS with $Q^2>1\gev^2$ using
(a) all proton candidates,
(b) proton candidates with momentum below $1.35\gev$ and $dE/dx$
above $1.3\,$mips, and
(c) proton candidates with momentum above $2\gev$.
The extended mass difference is defined as
$\Delta M^{\rm ext}=M(K \pi \pi_s p)-M(K \pi \pi_s)$ and
$M(\dsp)_{\rm PDG}$ is the nominal $\dsp$ mass.
The histograms  show the $M(D^* p)$
distributions for the like-sign combinations.
}
\label{fig:kpi_dp_dis}
\end{figure}

Figure~\ref{fig:kpi_dp_dis}a shows the $M(D^*p)$
distribution,
obtained with $D^*$ meson candidates reconstructed in the decay
channel (1),
for DIS with $Q^2>1\,$GeV$^2$.
Figures~\ref{fig:kpi_dp_dis}b and~\ref{fig:kpi_dp_dis}c
show the $M(D^*p)$ distributions
for the low-momentum and high-momentum proton selections,
respectively.
No narrow resonance is seen in either distribution.

\begin{figure}[hbtp]
\epsfysize=18cm
\vspace*{-1.0cm}
\centerline{\epsffile{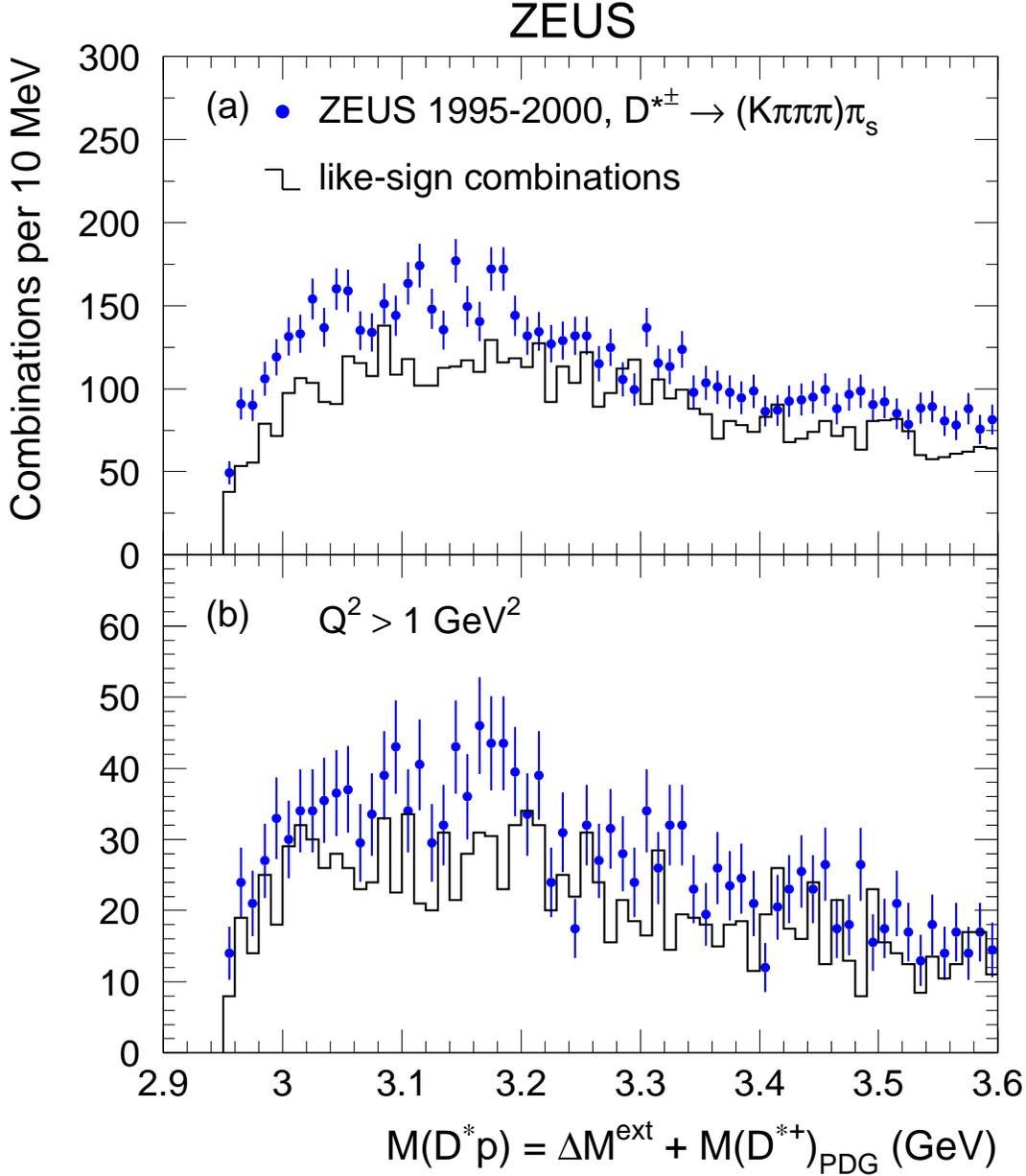}}
\caption{
The distribution of
$M(D^* p)=\Delta M^{\rm ext}+M(\dsp)_{\rm PDG}$
for charmed pentaquark candidates (dots)
(a) in the full data sample and
(b) in DIS with $Q^2>1\gev^2$.
The extended mass difference is defined as
$\Delta M^{\rm ext}=M(K \pi \pi \pi \pi_s p)-M(K \pi \pi \pi \pi_s)$ and
$M(\dsp)_{\rm PDG}$ is the nominal $\dsp$ mass.
The histograms  show the $M(D^* p)$
distributions for the like-sign combinations.
}
\label{fig:k3pi_dp}
\end{figure}

Figure~\ref{fig:k3pi_dp} shows the $M(D^*p)$
distributions,
obtained with $D^*$ meson candidates reconstructed in the decay
channel (2),
for the full data sample (Fig.~\ref{fig:k3pi_dp}a)
and for DIS with $Q^2>1\,$GeV$^2$ (Fig.~\ref{fig:k3pi_dp}b).
Both distributions show no narrow resonance.
No resonance was also observed using
the low-momentum and high-momentum proton selections
with $D^*$ meson candidates reconstructed in the decay channel (2) (not shown).

The histograms in Figs.~\ref{fig:kpi_dp_all}-\ref{fig:k3pi_dp} show the
$M(D^*p)$ distributions
for like-sign combinations
of $\dspm$ and proton candidates.
The shapes of the mass distributions for the unlike-sign and like-sign
combinations are similar.
The like-sign distributions lie
below the unlike-sign distributions at low $M(D^*p)$ values,
which is consistent with MC predictions.

\subsection{Systematic checks}
\label{sec-checks}

The selection cuts were varied to check
that the pentaquark signal was not lost due to some
specific selection requirement or hidden by the combinatorial background.
In particular,
the following systematic checks were carried out:
\begin{itemize}
\item variations were made in the cuts on $l_p$ and on the number
of CTD hits used for the $dE/dx$ measurement;
\item the cut on $l_p$ was replaced by a requirement for
proton candidate tracks to lie within
a wide $dE/dx$ band~\cite{pl:b591:7};
\item the high-momentum proton selection was repeated without cuts
on $l_p$
or on the number of
CTD hits used for the $dE/dx$ measurement;
\item to reduce the pion background in the proton candidate sample,
reflections from the decays of the excited $D$ mesons,
$\done,\dtwo\ra\dspm\pi^\mp$, to the $M(\dspm p^\mp)$ spectra were removed
by excluding all combinations with $2.38<M(\dspm \pi^\mp)<2.5\,$GeV;
\item DIS events were selected with $Q^2>20\,$GeV$^2$, i.e. in the range
where the cleanest $\Theta^+$ signal was observed in the previous
ZEUS analysis~\cite{pl:b591:7}.
Using this selection,
the numbers of reconstructed $D^*$ mesons were $2326\pm67$ in channel (1)
and $1799\pm78$ in channel (2);
\item DIS events were selected using only the inclusive DIS trigger.
Using this selection,
the numbers of reconstructed $D^*$ mesons were $3426\pm82$ in channel (1)
and $2369\pm86$ in channel (2);
\item tracks not assigned to the primary event vertex were
used together with the primary vertex tracks
for $D^*$ reconstruction and proton candidate selection.
\end{itemize}
No signal was observed using any of these selection variations.

The analysis was also repeated for the $D^*$ decay channel (1)
using very similar
selection criteria used in the analysis of
the H1 collaboration~\cite{pl:b588:17}.
The minimum transverse momentum requirements applied to tracks
forming $D^*$ combinations
were set to the H1 values.
The cut $p_T(D^*)/\etw10 > 0.12$
used in the ZEUS analysis
was replaced by the cut
$z(D^*)>0.2$, where
$z(D^*)= P\cdot p(D^*) / P\cdot q $ and
$P$, $p(D^*)$ and $q$ are the four-momenta of the incoming
proton, the $D^*$ meson and the exchanged photon.
In the proton rest frame, $z(D^*)$ is the fraction of the photon energy
carried by the $\dspm$ meson. The requirements on $M(K\pi)$ and $\Delta M$
were kept as in the nominal ZEUS analysis
since they were determined by the mass resolution of the ZEUS CTD.\footnote{
The check was also repeated with the H1 requirements on $M(K\pi)$
and $\Delta M$.}
The DIS events were selected with $Q^2>1\,$GeV$^2$ and $0.05<y<0.7$,
while
the photoproduction events were selected with $Q^2<1\,$GeV$^2$ and $0.2<y<0.8$.
The $D^*$ candidates were required to have
$-1.5<\eta(D^*)<1.0$ and
$p_T(D^*)>1.5\gev$ or $p_T(D^*)>2.0\gev$ in DIS or photoproduction
selections, respectively.
The numbers of reconstructed $D^*$ mesons
found using these cuts
were $5920\pm90$ and $11670\pm140$ for the DIS and photoproduction
selections, respectively.
To select proton candidates,
the requirement $l_p>0.15$ was replaced
by the H1 requirements on the normalised proton likelihood~\cite{pl:b588:17}.
The range of the proton momentum $1.6-2.0\,$GeV was excluded
in the case of photoproduction.

\begin{figure}[hbtp]
\epsfysize=18cm
\vspace*{-1.0cm}
\centerline{\epsffile{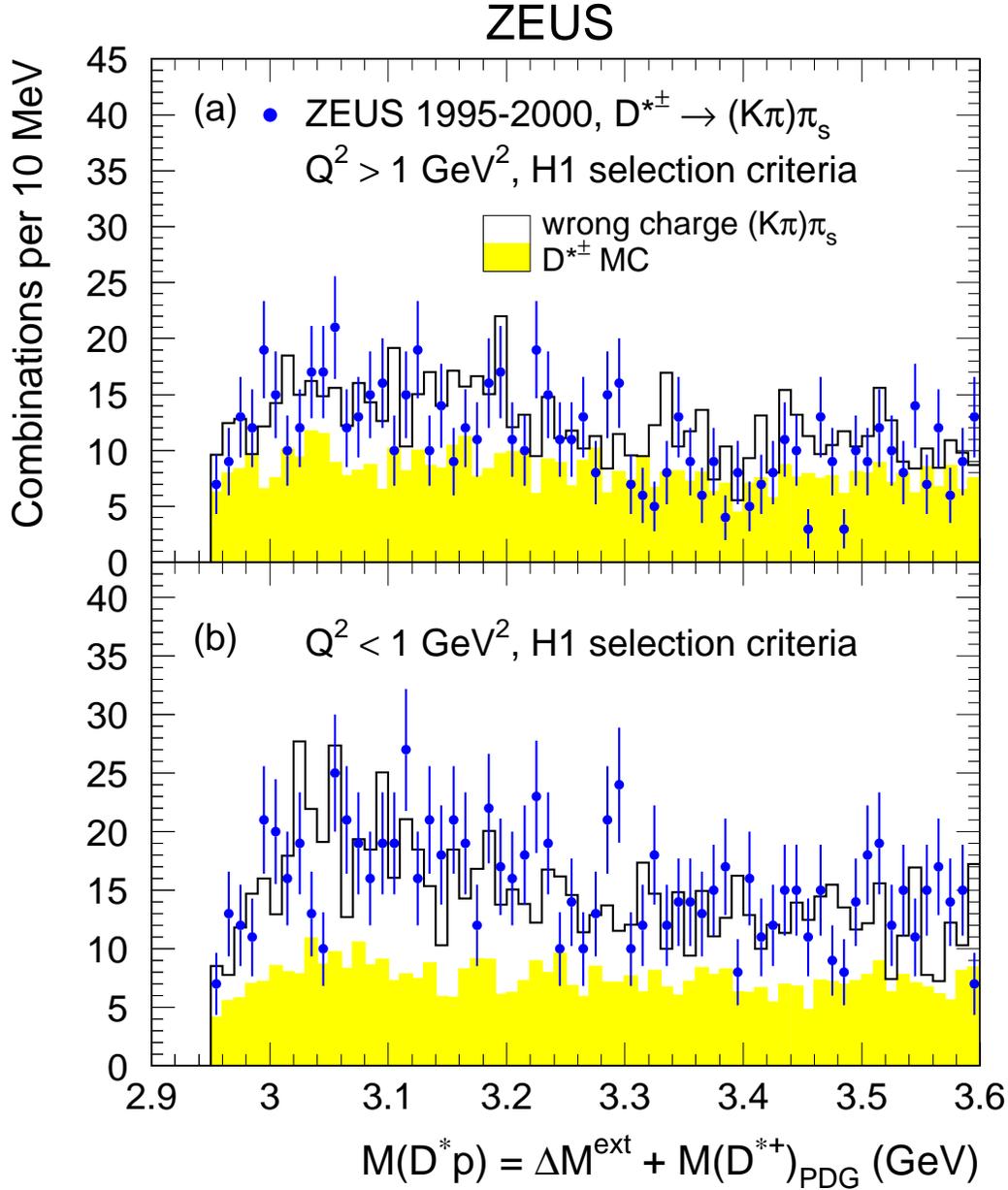}}
\caption{
The distribution of
$M(D^* p)=\Delta M^{\rm ext}+M(\dsp)_{\rm PDG}$
for charmed pentaquark candidates (dots)
obtained using H1 selection criteria in
(a) DIS with $Q^2>1\gev^2$
and (b) photoproduction with $Q^2<1\gev^2$.
The extended mass difference is defined as
$\Delta M^{\rm ext}=M(K \pi \pi_s p)-M(K \pi \pi_s)$ and
$M(\dsp)_{\rm PDG}$ is the nominal $\dsp$ mass.
The histograms show a two-component model in which
the wrong charge $(K\pi)\pi_s$ combinations are used to describe
the non-charm contribution and the inclusive $D^{*\pm}$ Monte Carlo
simulation (shaded area) describes the contribution of real $D^{*\pm}$ mesons.
}
\label{fig:kpi_h1}
\end{figure}

Figure~\ref{fig:kpi_h1} shows the $M(D^*p)$ distributions separately
for the DIS and photoproduction events selected using the H1 criteria.
There is no indication of a narrow resonance in either distribution.
Yields of combinations in the ZEUS and H1 $M(D^*p)$ distributions for DIS
are in approximate proportion to the
corresponding numbers
of $D^*$ mesons.
The histograms in Fig.~\ref{fig:kpi_h1}
show a two-component model in which
the wrong charge $(K\pi)\pi_s$ combinations,
normalised as described in Section 4,
were used to describe
the non-charm contribution, and the inclusive $D^*$ MC
simulation,
normalised to the $D^*$ yield in the data,
described the contribution of real $D^*$ mesons.
The model describes the measured $M(D^*p)$ distributions well.

\section{Evaluation of upper limits}
\label{sec-limits}

\begin{figure}[hbtp]
\epsfysize=18cm
\vspace*{-1.0cm}
\centerline{\epsffile{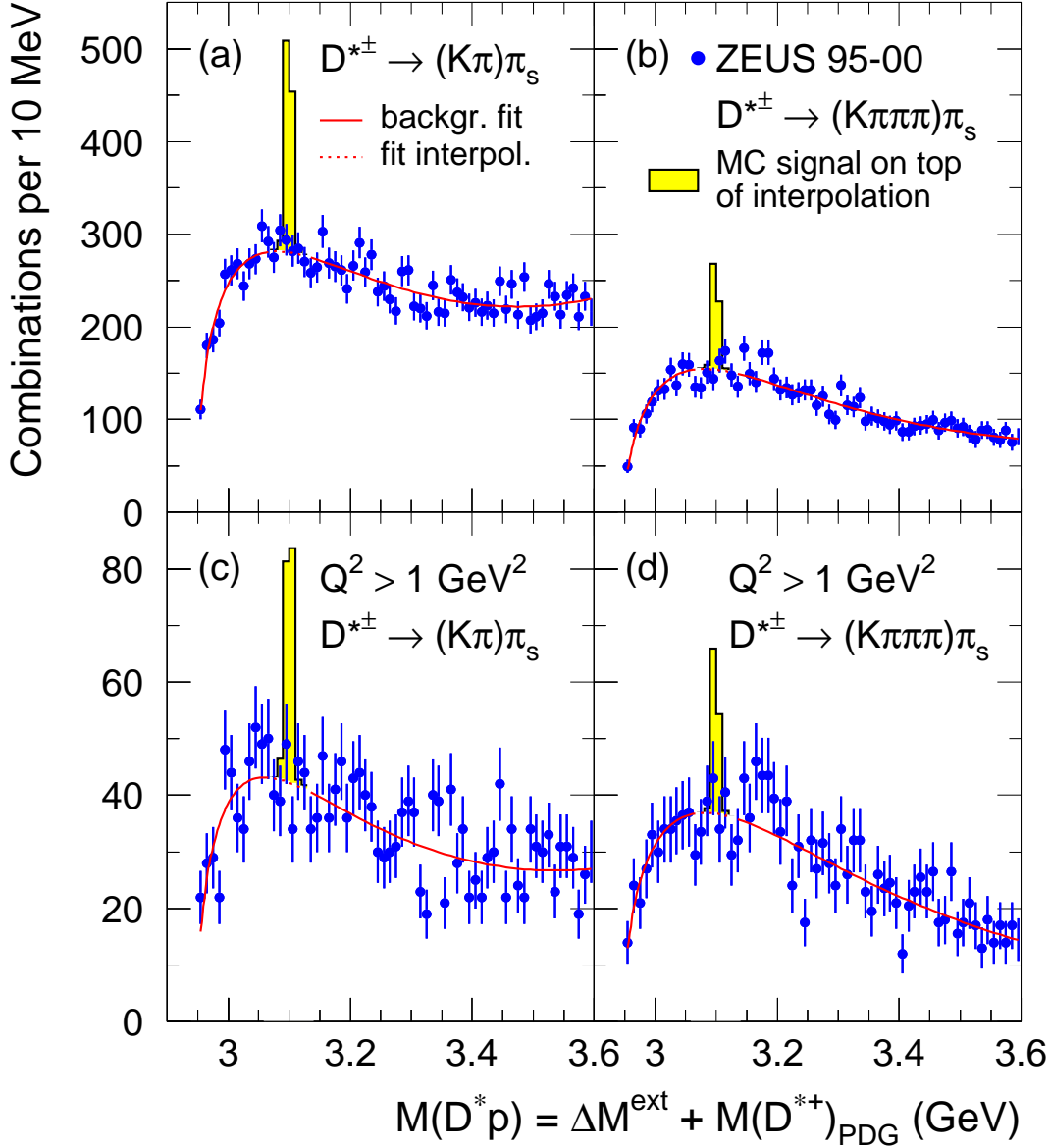}}
\caption{
The distribution of
$M(D^* p)$
for charmed pentaquark candidates (dots)
selected in
(a) the full data sample using
$D^{*\pm}\rightarrow (K\pi)\pi_s$ candidates,
(b) the full data sample using
$D^{*\pm}\rightarrow (K\pi\pi\pi)\pi_s$ candidates,
(c) DIS with $Q^2>1\gev^2$ using
$D^{*\pm}\rightarrow (K\pi)\pi_s$ candidates and
(d) DIS with $Q^2>1\gev^2$ using
$D^{*\pm}\rightarrow (K\pi\pi\pi)\pi_s$ candidates.
The solid curves are fits to the background
function outside the signal window $3.07-3.13\gev$.
The shaded histograms show the Monte Carlo $\Theta^0_c$ signals,
normalised to $1\%$ of the number of reconstructed $D^*$ mesons,
and shown on top of the fit interpolations (dashed curves) in the signal window.
}
\label{fig:kpi_k3pi_fit}
\end{figure}

Upper limits on the fraction of $D^*$ mesons originating
from the $\Theta^0_c$ decays were set in the signal window
$3.07<M(D^*p)<3.13\,$GeV.
This window covers the H1 measurement taking into account the uncertainties
of the measured $\Theta^0_c$ mass and width.
The upper limits were calculated for the full $D^*$-meson samples
obtained with $D^*$ reconstructed in channels (1) and (2),
see Figs.~\ref{fig:kpi_k3pi_fit}a and~\ref{fig:kpi_k3pi_fit}b.
The calculations were also separately repeated with the samples obtained
in DIS (see Figs.~\ref{fig:kpi_k3pi_fit}c and~\ref{fig:kpi_k3pi_fit}d)
and photoproduction (not shown).
Each $M(D^*p)$ distribution was fitted outside the signal window
to the functional form
$x^a \exp(-bx+cx^2)$, where $x=\Delta M^{\rm ext}-m_{p}$ and $m_{p}$
is the proton mass.
The fitted curves describe the
$M(D^*p)$ distributions reasonably well in the whole range shown
in Fig.~\ref{fig:kpi_k3pi_fit}.
The number of reconstructed $\Theta^0_c$ baryons was estimated
by subtracting the background function, integrated over the signal window,
from the observed number of candidates in the window.
This number was divided by the number of reconstructed
$D^*$ mesons, yielding
the fraction of $D^*$ mesons originating
from the $\Theta^0_c$ decays,
$R(\Theta^0_c\rightarrow D^*p/D^*)$.

The numbers used for the upper-limit calculations and
the measured upper limits are summarised in Table~1.
The reported upper limits are the frequentist confidence bounds
calculated assuming a Gaussian probability function in
the unified approach~\cite{pr:d57:3873}.
The results are shown separately for the full data sample, for DIS with
$Q^2>1\,$GeV$^2$ and for photoproduction with $Q^2<1\,$GeV$^2$.

The 95\% C.L. upper limits on
$R(\Theta^0_c\rightarrow D^*p/D^*)$
were found to be
0.29\% and 0.33\%
for the full $D^*$-meson samples
obtained with $D^*$ reconstructed in channels (1) and (2), respectively.
To average the $R(\Theta^0_c\rightarrow D^*p/D^*)$ values obtained
with $D^*$ reconstructed in the two decay channels,
a standard weighted least-square procedure~\cite{pr:d55:10001} was used.
The combined upper limit from both decay channels is 0.23\%.
The combined upper limit for DIS with $Q^2>1\gev^2$ is 0.35\%.

The H1 Collaboration reported a $\Theta^0_c$ baryon
contributing roughly $1\%$ of the $D^*$ production rate,
in the kinematic region studied in that analysis,
in DIS with $Q^2>1\gev^2$,
and a clear signal of compatible mass and width in a photoproduction
sample ($Q^2<1\gev^2$)~\cite{pl:b588:17}.
If the $\Theta^0_c$ baryon contributed $1\%$ of the number
of $D^*$ mesons in the kinematic region studied in this analysis,
a signal of $626$ $\Theta^0_c$ baryons
would be expected using the full samples of the $D^*$ mesons
reconstructed in both decay channels.
Assuming Gaussian statistics, such a signal together with the expected
number of background events could produce the observed number of events
in the signal window only in cases of
statistical fluctuations larger than $9\,\sigma$.
A production rate corresponding to $1\%$ of $D^*$'s of the present
analysis in the DIS ($Q^2>1\gev^2$) sample only
is excluded at $5\,\sigma$.
In Fig.~\ref{fig:kpi_k3pi_fit},
the MC $\Theta^0_c$ signals normalised to $1\%$ of the numbers of
reconstructed $D^*$ mesons are
shown on top of
the fitted backgrounds.

To correct the fraction
of $D^*$ mesons originating from the $\Theta^0_c$ decays for detector
effects,
the relative acceptance was calculated
from the acceptances defined in Sections 4 and 5 as:
$$\frac{A(\Theta^0_c\rightarrow D^*p)}{A^{\rm inc}(D^*)}=
\frac{A^{\Theta^0_c}(D^*)}{A^{\rm inc}(D^*)} \cdot A(p) \cdot A(l_p>0.15).$$
The systematic uncertainty of the background fit procedure
was estimated by varying the range used in the fit.
To estimate the systematic uncertainty in the MC correction
factors, the $p_T(\Theta^0_c)$ and $\eta(\Theta^0_c)$ spectra of
the $\Theta^0_c$ MC were varied.
Both systematic uncertainties and the statistical uncertainties
of the data, MC and $A(l_p>0.15)$  were added in quadrature to determine
the total uncertainty used for the upper-limit calculation.
The 95\% C.L. upper limits on the corrected fraction of $D^*$ mesons
originating
from $\Theta^0_c$ decays,
$R^{\rm cor}(\Theta^0_c\rightarrow D^*p/D^*)$,
were found to be
0.47\% and 0.50\%
for the full $D^*$-meson samples
obtained with $D^*$ reconstructed in channels (1) and (2), respectively.
The combined upper limit from both decay channels is
0.37\%.
The effect of correlated systematic uncertainties was negligible in
the combined upper limit calculation.

The product of the fraction of $c$ quarks hadronising
as a $\Theta^0_c$ baryon, $f(c\rightarrow\Theta^0_c)$, and the branching ratio
of the $\Theta^0_c$ decay to $D^*p$, $B_{\Theta^0_c\rightarrow D^*p}$,
can be calculated as:
$$f(c\rightarrow\Theta^0_c)\cdot B_{\Theta^0_c\rightarrow D^*p} =
\frac{N(\Theta^0_c\rightarrow D^*p)}{N(D^*)} \cdot \fcds,$$
where $\fcds$ is the known rate of $c$ quarks hadronising as
$D^{*+}$ mesons~\cite{hep-ex-9912064p} and the ratio of
the numbers of the $\Theta^0_c$
and $D^*$ hadrons,
$\frac{N(\Theta^0_c\rightarrow D^*p)}{N(D^*)}$,
is calculated in the full phase space.
An extrapolation of the fractions
measured in the restricted $p_T(D^*)$ and $\eta(D^*)$ kinematic ranges
to the full phase space
would require precise modelling of the
$p_T(\Theta^0_c)$ and $\eta(\Theta^0_c)$ spectra. Such modelling
is currently not available.
To estimate the upper limit on
$f(c\rightarrow\Theta^0_c)\cdot B_{\Theta^0_c\rightarrow D^*p}$,
the corrected fractions of $D^*$ mesons
originating from the $\Theta^0_c$ decays were converted to the ratios
of numbers of the $\Theta^0_c$ and $D^*$ hadrons in their
respective kinematic ranges
used for the $D^*$ meson selection:
$$
\frac{N(\Theta^0_c\rightarrow D^*p; p_T(\Theta^0_c)>1.35, 2.8\,{\rm GeV}; |\eta(\Theta^0_c)|<1.6)}{N(D^*; p_T(D^*)>1.35, 2.8\,{\rm GeV}; |\eta(D^*)|<1.6)}=
R^{\rm cor}(\Theta^0_c\rightarrow D^*p/D^*)\cdot
f^{\rm conv},
$$
$$
f^{\rm conv} =
\frac{N(\Theta^0_c\rightarrow D^*p; p_T(\Theta^0_c)>1.35, 2.8\,{\rm GeV}; |\eta(\Theta^0_c)|<1.6)}{N(\Theta^0_c\rightarrow D^*p; p_T(D^*)>1.35, 2.8\,{\rm GeV}; |\eta(D^*)|<1.6)}.
$$
The conversion factors, $f^{\rm conv}$, obtained with the $\Theta^0_c$ MC,
were 1.6 and 2.8 for $p_T>1.35\,$GeV and $p_T>2.8\,$GeV, respectively.
Using these conversion factors,
the 95\% C.L. upper limits on
$f(c\rightarrow\Theta^0_c)\cdot B_{\Theta^0_c\rightarrow D^*p}$,
were estimated to be 0.18\% and 0.33\% 
for the full $D^*$-meson samples
obtained with $D^*$ reconstructed in channels (1) and (2), respectively.
The combined upper limit from both decay channels is
0.16\%.
The effect of correlated systematic uncertainties was negligible in
the combined upper limit calculation.

\section{Summary}
\label{sec-conc}

A resonance search has been made in the $\dspm p^\mp$ invariant-mass
spectrum with the ZEUS detector at HERA using an 
integrated luminosity of $126\pbi$.
The decay channels
$\dsp\rightarrow\dz\pi^{+}_{s}\rightarrow(K^{-}\pi^{+})\pi^{+}_{s}$
and
$\dsp\rightarrow\dz\pi^{+}_{s}\rightarrow(K^{-}\pi^{+}\pi^{+}\pi^{-})\pi^{+}_{s}$
(and the corresponding antiparticle decays) 
were used to identify $\dspm$ mesons.
No resonance structure was observed in the $M(\dspm p^\mp)$ spectrum
from more than $60\,000$ reconstructed $\dspm$ mesons.
The upper limit on the fraction of $D^*$ mesons
originating
from $\Theta^0_c$ decays
is 0.23\% (95\% C.L.).
The upper limit for DIS with $Q^2>1\,$GeV$^2$ is 0.35\% (95\% C.L.).
Thus, the ZEUS data are not compatible with the H1 report of
$\Theta^0_c$ baryon production in DIS and photoproduction,
with a rate, in DIS, of roughly $1\%$
of the $D^*$ production rate.

\section{Acknowledgements}
\label{sec-ackn}

We would like to thank the DESY Directorate for their strong support
and encouragement. The remarkable achievements of the HERA machine
group were vital for the successful completion of this work
and are greatly appreciated.
We thank Marek Karliner, Harry Lipkin
and Torbj\"orn Sj\"ostrand
for useful discussions.

{
\def\bibname{\Large\bf References}
\def\refname{\Large\bf References}
\pagestyle{plain}
\ifzeusbst
  \bibliographystyle{./BiBTeX/bst/l4z_default}
\fi
\ifzdrftbst
  \bibliographystyle{./BiBTeX/bst/l4z_draft}
\fi
\ifzbstepj
  \bibliographystyle{./BiBTeX/bst/l4z_epj}
\fi
\ifzbstnp
  \bibliographystyle{./BiBTeX/bst/l4z_np}
\fi
\ifzbstpl
  \bibliographystyle{./BiBTeX/bst/l4z_pl}
\fi
{\raggedright
\bibliography{./BiBTeX/user/syn.bib,%
              ./BiBTeX/bib/l4z_articles.bib,%
              ./BiBTeX/bib/l4z_books.bib,%
              ./BiBTeX/bib/l4z_conferences.bib,%
              ./BiBTeX/bib/l4z_h1.bib,%
              ./BiBTeX/bib/l4z_misc.bib,%
              ./BiBTeX/bib/l4z_old.bib,%
              ./BiBTeX/bib/l4z_preprints.bib,%
              ./BiBTeX/bib/l4z_replaced.bib,%
              ./BiBTeX/bib/l4z_temporary.bib,%
              ./BiBTeX/bib/l4z_zeus.bib,%
              ./BiBTeX/user/charm5q.bib,%
              ./BiBTeX/user/chadr.bib,%
              ./BiBTeX/user/dstargamma.bib,%
              ./BiBTeX/user/eps497.bib}}
}
\vfill\eject

%
\begin{table}[hbt]
\begin{center}
\begin{tabular}{|c|c|c|c|} \hline
$D^*$ decay &
$(K\pi)\pi_s$ &
$(K\pi\pi\pi)\pi_s$ &
Both \\
channel &&& channels\\
\hline
\hline
\multicolumn{4}{|c|}{Full data sample}\\
\hline
\hline
$N_{\rm window}$ & 1710 & 914 & \\
\hline
$N_{\rm backgr}$ & $1678\pm 23$ &  $919\pm 19$ & \\
\hline
$N(D^*)$ & $42680\pm 350$ &  $19900\pm 250$ & \\
\hline
$R(\Theta^0_c\rightarrow D^*p/D^*)$
& $<0.29\%$ & $<0.33\%$ & $<0.23\%$ \\
\hline
$R^{\rm cor}(\Theta^0_c\rightarrow D^*p/D^*)$
& $<0.47\%$ & $<0.50\%$ & $<0.37\%$ \\
\hline
$f(c\rightarrow\Theta^0_c)\cdot B_{\Theta^0_c\rightarrow D^*p}$
& $<0.18\%$ & $<0.33\%$ & $<0.16\%$ \\
\hline
\hline
\multicolumn{4}{|c|}{DIS with $Q^2>1\gev^2$}\\
\hline
\hline
$N_{\rm window}$ & 252 & 220 & \\
\hline
$N_{\rm backgr}$ & $252.8\pm 9.2$ &  $219.8\pm 8.8$ & \\
\hline
$N(D^*)$ & $8680\pm 130$ &  $4830\pm 120$ & \\
\hline
$R(\Theta^0_c\rightarrow D^*p/D^*)$
& $<0.41\%$ & $<0.69\%$ & $<0.35\%$ \\
\hline
$R^{\rm cor}(\Theta^0_c\rightarrow D^*p/D^*)$
& $<0.59\%$ & $<1.06\%$ & $<0.51\%$ \\
\hline
$f(c\rightarrow\Theta^0_c)\cdot B_{\Theta^0_c\rightarrow D^*p}$
& $<0.20\%$ & $<0.56\%$ & $<0.19\%$ \\
\hline
\hline
\multicolumn{4}{|c|}{Photoproduction with $Q^2<1\gev^2$}\\
\hline
\hline
$N_{\rm window}$ & 1458 & 695 & \\
\hline
$N_{\rm backgr}$ & $1422\pm 21$ &  $694\pm 15$ & \\
\hline
$N(D^*)$ & $34000\pm 330$ &  $15070\pm 220$ & \\
\hline
$R(\Theta^0_c\rightarrow D^*p/D^*)$
& $<0.36\%$ & $<0.40\%$ & $<0.29\%$ \\
\hline
$R^{\rm cor}(\Theta^0_c\rightarrow D^*p/D^*)$
& $<0.60\%$ & $<0.60\%$ & $<0.47\%$ \\
\hline
$f(c\rightarrow\Theta^0_c)\cdot B_{\Theta^0_c\rightarrow D^*p}$
& $<0.23\%$ & $<0.43\%$ & $<0.21\%$ \\
\hline
\end{tabular}
\label{tab:lim_all}
\caption{
Numbers of the $M(D^*p)$ combinations in the signal window, $N_{\rm window}$;
fit background estimations, $N_{\rm backgr}$; numbers of reconstructed $D^*$
mesons, $N(D^*)$; $95\%\,$C.L. upper limits on
the uncorrected, $R(\Theta^0_c\rightarrow D^*p/D^*)$,
and corrected, $R^{\rm cor}(\Theta^0_c\rightarrow D^*p/D^*)$,
fractions
of $D^*$ mesons originating from $\Theta^0_c$ decays; and $95\%\,$C.L.
upper limits on the product of the fraction of $c$ quarks hadronising
as a $\Theta^0_c$ baryon, $f(c\rightarrow\Theta^0_c)$, and the branching ratio
of the $\Theta^0_c$ decay
to $D^*p$, $B_{\Theta^0_c\rightarrow D^*p}$.
The results are shown for the full data sample, for DIS with
$Q^2>1\gev^2$ and for photoproduction with $Q^2<1\gev^2$.
}
\end{center}
\end{table}

%
%
%
%
%
%
%
%
%
%
%
%
%
%
%

%
%
\end{document}